\documentclass[twocolumn,superscriptaddress,showpacs,nofootinbib,preprintnumbers,secnumarabic,amssymb, nobibnotes, aps, prd]{revtex4-2}
\usepackage[utf8]{inputenc}
\usepackage{graphicx}
\usepackage{latexsym,amsmath,amssymb,amsthm,lmodern,float,url,bbm}
\usepackage{bm}
\usepackage{natbib}
\usepackage{color}
\usepackage{microtype}
\usepackage{import}
\usepackage{bbold}
\usepackage[plain]{fancyref}
\usepackage{varioref}
\usepackage{slashed}
\usepackage{multirow}
\usepackage{tikz}
\usepackage{scrextend}
\usepackage{braket}
\usetikzlibrary{shapes}
\usetikzlibrary{positioning}
\usepackage[normalem]{ulem}
\usepackage{comment}
\usepackage{amsmath}

\newcommand{\fig}[1]{Fig.~\ref{fig:#1}}

\newcommand{\eq}[1]{Eq.~(\ref{eq:#1})}

\usepackage[colorlinks=true,backref=false, linktocpage=true,
citecolor=blue,urlcolor=blue,linkcolor=blue,pdfpagemode=UseOutlines]{hyperref}
\hypersetup{
  bookmarksnumbered=true,
  pdftitle = {},
  pdfsubject = {},
  pdfauthor = {},
  pdfkeywords = {}
}

\begin{document}
\title{Towards Quantum Simulations of Sphaleron Dynamics at Colliders} 
\author{Min Huang}
\email{huangmin@ihep.ac.cn}
\affiliation{Theoretical Physics Division, Institute of High Energy Physics, Chinese Academy of Sciences, Beijing 100049, China}
\affiliation{School of Physics, University of Chinese Academy of Sciences, Beijing 100049, China}
\author{Ying-Ying Li}
\email{liyingying@ihep.ac.cn}
\affiliation{Theoretical Physics Division, Institute of High Energy Physics, Chinese Academy of Sciences, Beijing 100049, China}
\author{Yandong Liu}
\email{ydliu@bnu.edu.cn}
\affiliation{Key Laboratory of Beam Technology of Ministry of Education, School of Physics and Astronomy, Beijing Normal University, Beijing, 100875, China}
\affiliation{Institute of Radiation Technology, Beijing Academy of Science and Technology, Beijing 100875, China}
\author{Hao Zhang}
\email{zhanghao@ihep.ac.cn}
\affiliation{Theoretical Physics Division, Institute of High Energy Physics, Chinese Academy of Sciences, Beijing 100049, China}
\affiliation{School of Physics, University of Chinese Academy of Sciences, Beijing 100049, China}
\affiliation{Center for High Energy Physics, Peking University, Beijing 100871, China}

\date{\today}

\begin{abstract}
Sphaleron dynamics in the Standard Model at high-energy particle collisions remains experimentally unobserved, with theoretical predictions hindered by its nonperturbative real-time nature. In this work, we investigate a quantum simulation approach to this challenge. Taking the $1+1$D $O(3)$ model as a protocol towards studying dynamics of sphaleron in the electroweak theory, we identify the sphaleron configuration and establish lattice parameters that reproduce continuum sphaleron energies with controlled precision. We then develop quantum algorithms to simulate sphaleron evolutions where quantum effects can be included. This work lays the ground to establish quantum simulations for studying the interaction between classical topological objects and particles in the quantum field theory that are usually inaccessible to classical methods and computations.
\end{abstract}

\maketitle

\section{Introduction}
As a saddle-point of finite energy in the field space, sphaleron is a static physical solution to the electroweak field equations in Minkowski spacetime \cite{Simms_1964, Taubes:1982ie,Manton:1983nd,Forgacs:1983yu,Klinkhamer:1984di}. In the standard model (SM), the sphaleron process leads to baryon and lepton number violations, which at high temperature, can be a key ingredient for baryogenesis that aim to explain the origin of the observed baryon asymmetry \cite{Sakharov:1967dj, cohen_progress_1993}, see recent reviews \cite{Morrissey:2012db}. The sphaleron rates at finite temperature can be calculated using large-scale lattice simulations \cite{Ambjorn:1990pu,DOnofrio:2014rug}.
Additionally, there has been longstanding efforts in investigating whether such processes could be detected in high-energy collisions, such as those predicting the sphaleron production rates at colliders \cite{Rubakov:1996vz, Gibbs_1995, Bezrukov:2003qm, Bezrukov:2003er, Ringwald_2003, Ringwald_2003-1, Tye:2015tva, Tye:2017hfv}, and phenemenological studies of baryon number-violating processes at LHC \cite{Ellis:2016ast} or ultra-high-energy neutrino collisions \cite{Ellis:2016dgb}. Yet, the production rate of sphalerons in high-energy collisions has been calculated using various methods, resulting in predictions that differ by orders of magnitude \cite{Tye:2015tva}. Such large discrepancies might be due to the real-time nature of multisphaleron processes that calculations in Euclidean spacetime fail to capture \cite{Tye:2017hfv}. Thus, it is warranted to perform a renewed study of baryon number-violating processes in high-energy collisions based on real-time simulations achieved using the Hamiltonian descriptions.

In such collisions, incoming and outgoing particles correspond to excitations of quantum fields. Hamiltonians describing dynamics of these fields require reducing their infinite degrees of freedom to finite ones. By discretizing continuous spacetime into lattices and truncating local field amplitudes to finite values, the resulting Hilbert space, although finite in size, still grows exponentially with increasing spatial volume. With the advances in quantum technology and the emergence of a new computing paradigm, namely quantum computing, the number of storage units (qubits) is expected to grow polynomially with the system size, bringing new opportunities for performing such calculations non-perturbatively. Simulating dynamics in quantum field theories with quantum computer has also seen rapid advances in recent years, see through reviews \cite{PRXQuantum.4.027001, DiMeglio:2023nsa, np-review, fang2025quantumfrontiershighenergy}.

However, to achieve the goal of simulating a sphaleron process in particle collisions requires investigations into a line of questions, $\textit{e.g.}$, what are instanton solutions like on the lattice, how to manifest the sphaleron configurations based on a Hamiltonian with second quantizations. In this work, we initiate a first step towards this goal, which is investigating the sphaleron configuration on lattices. We will focus on the 1+1d $O(3)$ for its remarkable similarities to non-abelian gauge theories \cite{Novikov:1984ac}. With a $O(3)$ symmetry breaking term, this model admits the existance of sphaleron configuration, which can potentially serve as a toy model to investigate sphaleron productions at high energy collisions in the EW theory. The paper is organized as follows. In Sec. \ref{sec:2}, we give a quick review of the sphaleron in a $O(3)$ nonlinear $\sigma$-model. In Sec. \ref{sec:3}, we discuss the sphaleron configuration on lattices. Quantum simulation of sphaleron evolution will be outlined in Sec.\ref{sec:4}. We then conclude in Sec.\ref{sec:5}. 

\section{The nonlinear sigma model} 
\label{sec:2}
Simulating the sphaleron process in Weinberg-Salam theory is beyond the capability of quantum devices in the Noisy Intermediate-Scale Quantum (NISQ) era, where $\mathcal{O}(100)$ qubits without error corrections are available. Fortunately, the sphaleron process exists widely in a lot of field theory models when there is uncontractible loop in the configuration space of the classical field \cite{Forgacs:1983yu}. The $O(3)$ nonlinear $\sigma$-model in $1 + 1$ dimensional (1+1D) spacetime, has been studied before for its remarkable similarities to non-abelian gauge theories \cite{Novikov:1984ac}. When modified by an explicit symmetry-breaking term that breaks the $O(3)$ symmetry down to an $O(2)$ subgroup \cite{Mottola:1988ff}, it could be presented as a model for baryon-number and lepton-number violation which are essential ingredient to explain the matter-antimatter asymmetry observed in our universe. Here we will investigate the sphaleron dynamics in this model with quantum computer as a first effort towards to quantum simulate the sphaleron process  in the Weinberg-Salam theory.

The classical action of the $O(3)$ nonlinear $\sigma$-model with explicit symmetry-breaking term in 1+1D spacetime is (with $\omega>0$)
\begin{equation}
S=\frac{1}{g^2}\int {\mathrm{d}}t~{\mathrm{d}}x\left[\frac{1}{2}\partial_\mu{\mathbf n}\cdot\partial^\mu{\mathbf n}-\omega^2(1+{\mathbf n}\cdot\mathbf{e}_z)\right],
  \label{eq:origin}
\end{equation}
where ${\mathbf n}(x)\cdot{\mathbf n}(x)\equiv 1$ is the normalized classical field from $\mathbbm R^2$ to $S^2$, $\mathbf{e}_z$ is the unit vector along the 3rd axis in the target space. The second term in the action explicitly breaks the global $O(3)$ symmetry to its $O(2)$ subgroup of the rotations around the 3rd axis in the target space. 

The field $\mathbf{n}$ can be parameterized as\footnote{Due to the nontrivial topology of $S^2$, this coordinate system has some singularities. So strictly speaking, it can be only safely used in the region of  $\eta\in(0,\pi)$, $\xi\in(0,2\pi)$.}
\begin{equation}
  {\mathbf n} (\xi, \eta) =\left(\begin{matrix}
  \sin \eta \sin \xi\\
  \sin \eta \cos \eta (\cos \xi - 1)\\
  -\sin^2 \eta \cos \xi - \cos^2 \eta
  \end{matrix}\right),
  \label{eq:nvec}
\end{equation}
and the action as a functional of the fields $\eta(t,x)$ and $\xi(t,x)$ can be written as
\begin{eqnarray}
    S&=&\frac{1}{g^2}\int {\mathrm{d}}t~{\mathrm{d}}x\biggl\{\frac{1}{2}\left[(1-\cos\xi)^2+\sin^2\xi\cos^2\eta\right]\partial_\mu\eta\partial^\mu\eta\nonumber\\
    &&+\sin\xi\sin\eta\cos\eta\partial_\mu\eta\partial^\mu\xi+\frac{1}{2}\sin^2\eta\partial_\mu\xi\partial^\mu\xi\nonumber\\
    &&-\omega^2\sin^2\eta(1-\cos\xi)\biggr\}.
\end{eqnarray}
The Hamiltonian for static configurations by neglecting the kinetic term as we derived in App.~(\ref{sec:app}) is 
\begin{eqnarray}
    H&=&\frac{1}{g^2}\int {\mathrm d}x\biggl[\frac{1}{2}\left(1-\cos\xi\right)^2\left(\frac{\partial \eta}{\partial x}\right)^2+\frac{1}{2}\biggl(\frac{\partial \eta}{\partial x}\sin\xi\cos\eta\nonumber\\
    &&+\frac{\partial \xi}{\partial x}\sin\eta\biggr)^2+\omega^2\sin^2\eta(1-\cos\xi)\biggr].
\end{eqnarray}
To find the sphaleron solution to the theory in the continuum, one considers static configurations $\hat{n}$ with fixed $\eta$, the energy obtained can be written as \cite{Mottola:1988ff}
\begin{equation}
E = \frac{\sin^2 \eta}{g^2} \int {\mathrm{d}} x \left[
   \frac{1}{2} \left( \frac{\partial \xi}{\partial x} \right)^2 + 
   \omega^2 (1 - \cos \xi) \right].
\end{equation}
As seen, $\eta = \pi/2$ gives the maximal of the energy functional. At this maximal, one solves the configuration of $\xi$ which minimizes the energy $E$ to find a saddle point. Notice that $\xi \equiv 2k\pi$ ($k\in\mathbbm Z$) corresponds to the ground state satisfying $\mathbf n(x)\equiv (0,0,-1)^{\mathrm{T}}$, independent of the $\eta$ values, which should not be considered as the sphaleron solutions. With the boundary condition\footnote{The sphaleron configuration is homotopically equivalent to the ground state configuration since $\pi_1(S^2)$ is trivial. One can verifies this conclusion with the solution (\ref{eq:sphaleron}) directly by transforming the field configuration along the path
\begin{equation}
\mathbf{n}(\xi(x,s),\eta(x,s))=\mathbf{n}\left(\xi^{\mathrm{sph}}(x),\frac{\pi}{2}(1-s)\right).
\end{equation}
It is easy to verify that this continuous map connects the sphaleron configuration ($s=0$) and the ground state configuration ($s=1$). The reason for such a boundary condition is because that the variation along the $\eta$-direction has been frozen. }
\begin{eqnarray}
    \lim_{x\to -\infty} \xi(x) = 0,  ~~~~\lim_{x\to+\infty} \xi(x) = 2\pi,
    \label{eq:boundary}
\end{eqnarray} 
the configuration of $\xi$ satisfying the Euler-Lagrange equation is solved as (since the arcsin function is not continued at $[0,2\pi]$)
\begin{equation}
\xi^{\rm sph} (x) = 
    \left\{
    \begin{array}{cc}
    2 \arcsin({\rm sech}(\omega x)),&x<0\\
    2\pi-2 \arcsin({\rm sech}(\omega x)),&x\geqslant0
    \end{array}
    \right.    ,    
\label{eq:sphaleron}
\end{equation}
or
\begin{equation}
  {\mathbf n} (x) =\left(\begin{matrix}
  -2{\mathrm{sech}}(\omega x)\tanh(\omega x)\\
  0\\
  -1+2{\mathrm{sech}}^2(\omega x)
  \end{matrix}\right).
\end{equation}
Substituting into the energy equation with $\eta = \pi/2$, one can find out the saddle point and get the sphaleron energy
\begin{equation}
  E_{\rm sph} \bigg|_{\eta = \frac{\pi}{2}} = \frac{8 \omega}{g^2}.
\end{equation}

To prove that Eq. (\ref{eq:sphaleron}) is a saddle point in the configuration space mathematically rigorously, we consider the infinitesimal variation
\begin{equation}
\xi(x)\to\xi^{\mathrm{sph}}(x)+\varepsilon\delta\xi(x),~\eta(x)\to \frac{\pi}{2}+\varepsilon\delta\eta(x),~\varepsilon\to 0^+.
\end{equation}
Then the variation of the Hamiltonian can be expanded to $\mathcal{O}(\varepsilon^2)$ as
\begin{equation}
H\to H^{(0)}+\varepsilon H^{(1)}+\varepsilon^2 H^{(2)}+\mathcal{O}(\varepsilon^3).
\end{equation}
The zeroth order term is just the Hamiltonian of Eq. (\ref{eq:sphaleron})
\begin{equation}
H^{(0)}=\frac{1}{g^2} \int {\mathrm{d}} x \left[
   \frac{1}{2} \left( \frac{\partial \xi^{\mathrm{sph}}}{\partial x} \right)^2 + 
   \omega^2 (1 - \cos \xi^{\mathrm{sph}}) \right]=\frac{8\omega}{g^2}.
\end{equation}
The linear term
\begin{eqnarray}
H^{(1)}&=&\frac{1}{g^2}\int {\mathrm{d}} x \biggl[\left( \frac{\partial \xi^{\mathrm{sph}}}{\partial x} \right)\left( \frac{\partial \delta\xi}{\partial x} \right)+\omega^2\delta\xi\sin\xi^{\mathrm{sph}}\biggr]\nonumber\\
&=&\frac{1}{g^2}\int {\mathrm{d}} x \biggl[\left( \frac{\partial \xi^{\mathrm{sph}}}{\partial x} \right)\left( \frac{\partial \delta\xi}{\partial x} \right)+\delta\xi\frac{\partial^2 \xi^{\mathrm{sph}}}{\partial x^2}\biggr]\nonumber\\
&=&\frac{1}{g^2}\int {\mathrm{d}} x \frac{\partial}{\partial x}\biggl[\delta\xi\frac{\partial \xi^{\mathrm{sph}}}{\partial x}\biggr]\nonumber\\
&=&\frac{1}{g^2} \delta\xi\frac{\partial \xi^{\mathrm{sph}}}{\partial x}\biggr|_{x\to-\infty}^{x\to+\infty}\nonumber\\
&=&\frac{2\omega}{g^2}\delta\xi~{\mathrm{sech}}(\omega x)\biggr|_{-\infty}^{+\infty}
\end{eqnarray}
since $\mathrm{sech}(\omega x)$ goes to 0 faster than any polynomial of $x$ when $x$ goes to infinity, we have $H^{(1)}=0$ if $\delta\xi$ is at most polynomially divergent when $x$ goes to infinity\footnote{If one defines the ``sphaleron'' solution as
\begin{equation}
    \xi^{\rm sph} (x) = 2 \arcsin({\rm sech}(\omega x)), 
    \label{eq:wrongsphaleron}
\end{equation}
without carefully treating of the discontinuity of the arcsin function, the linear term will become
\begin{eqnarray}
H^{(1)}&=&\frac{2\omega}{g^2}\int {\mathrm{d}} x \sqrt{\tanh^2(\omega x)}\left[-\left( \frac{\partial \delta\xi}{\partial x} \right){\mathrm{csch}}(\omega x)+\omega\delta\xi~{\mathrm{sech}}(\omega x)\right]\nonumber\\
&=&\frac{2\omega}{g^2}\int {\mathrm{d}} x \biggl[ -\left( \frac{\partial \delta\xi}{\partial x} \right){\mathrm{sech}}(\omega x)+\delta\xi\frac{\partial}{\partial x}\left(-{\mathrm{sech}}(\omega x)\right)\biggr]{\mathrm{sign}}(x)\nonumber\\
&=&\frac{2\omega}{g^2}\biggl\{\int_0^{+\infty} {\mathrm{d}} x \frac{\partial}{\partial x}\left[-\delta\xi~{\mathrm{sech}}(\omega x)\right]-\int_{-\infty}^0 {\mathrm{d}} x \frac{\partial}{\partial x}[-\delta\xi\nonumber\\
&&{\mathrm{sech}}(\omega x)]\biggr\}\nonumber\\
&=&-\frac{2\omega}{g^2} \left[\delta\xi~{\mathrm{sech}}(\omega x)_{x\to +\infty}+\delta\xi~{\mathrm{sech}}(\omega x)_{x\to -\infty}-2\delta\xi(0)\right],\nonumber\\
\end{eqnarray}
which is no longer a pure boundary term and not zero. Such a non-vanished first order variation is clearly a proof that the Eq. (\ref{eq:wrongsphaleron}) is not a static solution of the equation of motion, and could not be a sphaleron.}.
\begin{figure}
    \centering    \includegraphics[width=0.8\linewidth]{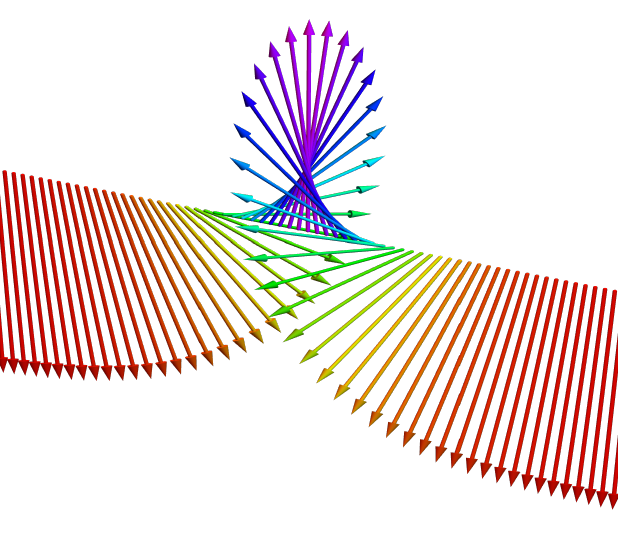}
    \caption{${\mathbf n}(x)$ configurations in the $x-z$ plane at different $x$ for the sphaleron solution. }
    \label{fig:true_sphaleron}
\end{figure}
\begin{figure}
    \centering    \includegraphics[width=0.8\linewidth]{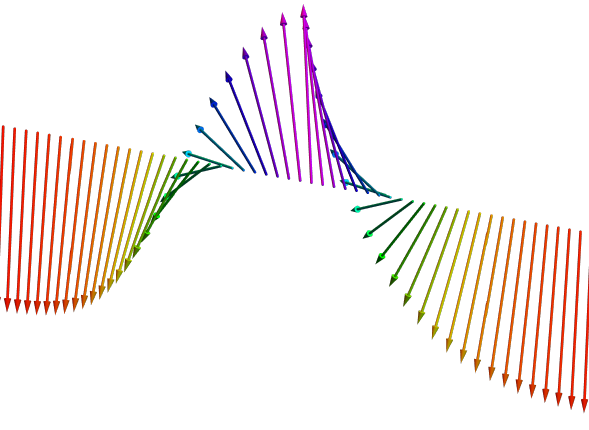}
    \caption{Same as FIG. \fig{true_sphaleron} but for the solution in Eq.(\ref{eq:wrongsphaleron}). }
    \label{fig:false_sphaleron}
\end{figure}

The quadratic term can be expanded as
\begin{eqnarray}
    H^{(2)}&=&\frac{1}{2g^2}\int {\mathrm{d}} x\biggl[\omega^2(1-2{\mathrm{sech}}^2(\omega x))\delta\xi^2+\left(\frac{\partial\delta\xi}{\partial x}\right)^2\nonumber\\
    &&-8\omega^2{\mathrm{sech}}^2(\omega x)\delta\eta^2+\frac{8\omega\sinh(\omega x)}{\cosh^3(\omega x)}\delta\eta\left(\frac{\partial \delta\eta}{\partial x}\right)\nonumber\\
    &&+4{\mathrm{sech}}^4(\omega x)\left(\frac{\partial \delta\eta}{\partial x}\right)^2\biggr].
\end{eqnarray}
It is easy to verify that for the variation
\begin{equation}
\delta\xi=0,~\delta\eta=\theta(x_0-x)\theta(x_0+x),~x_0>0
\end{equation}
where $\theta(\cdot)$ is the Heaviside-$\theta$ function, the $H^{(2)}=-8\omega\tanh(\omega x_0)/g^2$ is negative. On the other hand, for the variation
\begin{equation}
\delta\xi=\theta(x_0-x)\theta(x_0+x),~\delta\eta=0,~x_0>0,
\end{equation}
we have
\begin{equation}
    H^{(2)}=\frac{\omega}{g^2}\left[\omega x_0-2{\mathrm{tanh}}(\omega x_0)\right],
\end{equation}
which is positive if $x_0>2/\omega$. Thus the second order variation of the energy of the static solution Eq. (\ref{eq:sphaleron}) is indefinite which means that the solution is a saddle point in the configuration space.

\section{Sphaleron Solutions on a Lattice}
\label{sec:3}
To simulate the two-dimensional $O(3)$ model with quantum algorithms, one first discretizes the space on lattices. The lattice Hamiltonian describing the $O (3)$ non-linear $\sigma$-model for a system with $2N+1$ sites and lattice spacing $a \equiv L/2N$, as we also derive in the App.~\ref{sec:app}, is given by \cite{Hamer:1978ew, Alexandru_2019, Alexandru_2023, Araz:2022tbd, Bruckmann:2018usp}
\begin{eqnarray}
    H^{\rm latt}_{\rm tot} &=& \frac{g^2}{2 a} \sum_{k = -N}^{N}
   \mathbf{L}^2_k + \frac{1}{a g^2} \sum_{k = -N}^{N - 1}
   (\mathbb{1} - {\mathbf n}_k \cdot {\mathbf n}_{k + 1}) \nonumber\\
   && + \frac{a \omega^2}{g^2} \sum_{k = -N}^{N} (1 + {\mathbf n}_k \cdot \mathbf{e}_z),
   \label{eq:quant-ham}
\end{eqnarray}
where the integer $k$ labels the spatial lattice and $a$ is the lattice spacing, respectively. The first term $\mathbf{L}^2_k$ is recognized as the angular momentum and the term ${\mathbf n}_k \cdot {\mathbf n}_{k + 1}$ as the interaction term between each lattice site and its nearest neighbors. As the sphaleron solution is a static solution where the field is time-independent, the angular momentum $\mathbf{L}^2_k$ could be neglected, which lead to the Hamiltonian:
\begin{equation}
    H^{\rm latt} = \frac{1}{ag^2} \underset{k = -N}{\overset{N - 1}{\sum}}
   (1-{{\mathbf n}}_k \cdot {{\mathbf n}}_{k + 1}) + \frac{a \omega^2}{g^2} \underset{k =
   -N}{\overset{N}{\sum}} (1+{\mathbf n}_k \cdot \mathbf{e}_z).
   \label{eq:H}
\end{equation}
for solving the sphaleron configurations on the lattice.

To study the sphaleron solution after discretization, it's necessary to revisit the coordinate system on the lattice at site $k$:
\begin{equation}
\label{eq:nvec}
  {\mathbf n}_k =\left(\begin{matrix}
  \sin \eta_k \sin \xi_k\\
  \sin \eta_k \cos \eta_k (\cos \xi_k - 1)\\
  -\sin^2 \eta_k \cos \xi_k - \cos^2 \eta_k
  \end{matrix}\right).
\end{equation}
The energy for fixed value of $\eta_k \equiv \eta_0$ can be written as
\begin{eqnarray}
 H^{\rm latt} &=&\frac{\sin^2 \eta_0}{ag^2}\biggl\{\underset{k =-N}{\overset{N-1}{\sum}}[1-\cos (\xi_k -
  \xi_{k + 1})]
  \nonumber\\
  &&~~~~~~~~~~~~~~~+a^2\omega^2\underset{k =-N}{\overset{N}{\sum}}(1- \cos \xi_k )\biggr\}.
  \label{eq:latt-energy}
\end{eqnarray}
To find the sphaleron configuration in the framework of the lattice field theory, we add the following boundary conditions along with $\eta_0=\pi/2$, as inspired by the continuous case:
\begin{equation}
    \xi_{-N}=0,~~\xi_N=2\pi,
    \label{eq:boundary-lattice1}
\end{equation}
and find the sequence $\{\xi_k\}$ which minimizes $H^{\rm latt}$. However, there is a fundamental difference between the continuous case and the discrete case since the coordinate space is no longer path connected, and it is meaningless to distinguish $0$ from $2\pi$ (for more discussions, see \cite{Chen:2024ddr} for example). This subtlety can be manifested by noticing that 
\begin{equation}
    \xi_k=\left\{
    \begin{array}{cc}
    0,&k<m\\
    2\pi,&k\geqslant m
    \end{array}
    \right.    ,    
\end{equation}
for any integer $m$ in the range $-N<m\leqslant N$ is always a (trivial) solution which minimizes the $H^{\rm latt}$ and satisfies the boundary condition (\ref{eq:boundary-lattice1}).

To resolve this subtlety and avoid the above trivial solution, we require at least one site $i$ with $\xi_i$ different from $0$ and $2\pi$. In the continuous case, when $x$ varies from $-\infty$ to $+\infty$, $\xi(x)$ takes on every value in the interval $[0,2\pi]$. So we can assign $\xi_i$ at site $i$ to $\alpha$ which is not integer times to $2\pi$. In principle, the different choice of $i$ and $\alpha$ corresponds the translation of the position of the sphaleron along the space direction. For simplicity and to minimize the boundary effect, we assign the site $i=0$ to have $\xi_0 = \alpha$, which leads to the condition:
\begin{equation}
    \xi_{-N}=0,~~\xi_0=\alpha,~~\xi_N=2\pi.
    \label{eq:boundary-lattice2}
\end{equation}
It is easy to see that the sequence which minimizes $H^{\rm latt}$ must have form
\begin{equation}
    0,~\xi_{N-1},~\cdots,~\xi_{1},~\alpha,~2\pi-\xi_1,~\cdots,~2\pi-\xi_{N-1},~2\pi.
    \label{eq:latt_seq1}
\end{equation}
Again, as we have pointed out, it is meaningless to distinguish this configuration from
\begin{equation}
    0,~\xi_{N-1},~\cdots,~\xi_{1},~\alpha,~\xi_1,~\cdots,~\xi_{N-1},~0
    \label{eq:latt_seq2}
\end{equation}
on a lattice. So we usually work on the second one for simplicity. 

We now want to show that the configuration (\ref{eq:latt_seq1}) goes to the classical sphaleron solution in the continuum limit. The proof will be seperated to two steps. First, we show that the consequence is monotonic. To show that, let us focus on the left half-segment: $\{0,\xi_{N-1},\cdots,\alpha\}$. If there exists three-site $\{\xi_{k-1},~\xi_k,~\xi_{k+1}\}$ which is not monotonic, then either
\begin{equation}
    {\text{Case I:}}~~\xi_{k-1}>\xi_k,~\xi_{k+1}>\xi_{k},
\end{equation}
or
\begin{equation}
    {\text{Case II:}}~~\xi_{k-1}<\xi_k,~\xi_{k+1}<\xi_{k}.
\end{equation}
The variation of the Hamiltonian to the variable $\xi_k$ is
\begin{eqnarray}
        \delta H^{\rm latt}&=&\frac{\sin^2\eta_0}{ag^2}[\sin(\xi_k-\xi_{k+1})+\sin(\xi_k-\xi_{k-1})\nonumber\\        &&+a^2\omega^2\sin\xi_k]\delta\xi_k.
\end{eqnarray}
In the continuum limit ($N\to+\infty,a=L/2N\to 0$ with $\omega$ and $g$ fixed), the third term vanishes as $\mathcal{O}(a^2)$, decaying faster than the first two terms which scale as $\mathcal{O}(a)$. So for Case I, $\delta H^{\rm latt}$ is negative when $\delta\xi_k>0$, and for Case II, $\delta H^{\rm latt}$ is negative when $\delta\xi_k<0$. Both these variations pull $\{\xi_{k-1},~\xi_k,~\xi_{k+1}\}$ back to a monotonic consequence.

Second, we show that the value of  $\alpha$ should be $\pi$. Let us consider the variation to the variable $\alpha$. 
\begin{equation}
        \delta H^{\rm latt}=\frac{\sin^2\eta_0}{ag^2}[2\sin(\alpha-\xi_1)+a^2\omega^2\sin\alpha]\delta\alpha.
\end{equation}
Once again, one may neglect the second term in the continuum limit since it is of higher order of $a$. Then if $\alpha>0$, the variation is greater than $0$ when $\delta\alpha>0$ since the sequence is increased in this case. If $\alpha<0$, the variation is greater than $0$ when $\delta\alpha<0$ since the sequence is increased. So to get the extreme value of the $H^{\rm latt}$, one has to push $\alpha$ to the boundary of its domain, which means that one should choose $\alpha=\pm\pi$. We would like to point it out that even with the constraints, the lattice model could not distinguish the sphaleron solution Eq. (\ref{eq:sphaleron}) from the ``wrong'' sphaleron solution Eq. (\ref{eq:wrongsphaleron}). One need additional conditions such as requiring $\xi(x)$ increasing from $0$ to $2\pi$ after taking the continuum limit. This is again because that the discrete model can not faithfully reflect the topological properties of the configuration \cite{Chen:2024ddr}. In this work, we add the constraints
\begin{eqnarray}
    &&\xi_{-N}=0,~~\xi_0=\pi,~~\xi_N=2\pi,\nonumber\\
    &&{\text{for }}\forall~k\in[-N,N),~~\xi_k<\xi_{k+1}.
\label{eq:boundaryeta}
\end{eqnarray}
The last constraint will guarantee the continuum limit of the configuration from the lattice model is Eq. (\ref{eq:sphaleron}) instead of Eq. (\ref{eq:wrongsphaleron}).

\begin{figure}[h]
\centering
    \includegraphics[width=0.45\textwidth,height=0.3\textwidth]{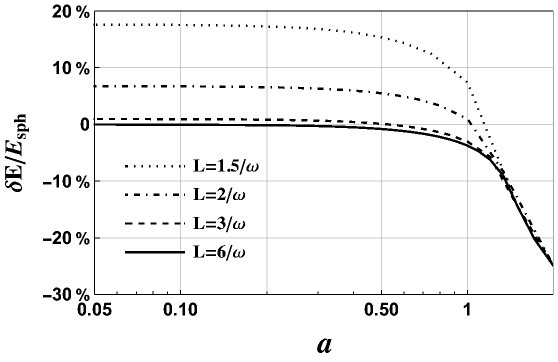}
    \caption{Relative errors of the sphaleron energy on the lattice to its continuum value $E_{\rm sph}$, with $\delta E\equiv E_{\mathrm{sph}}(L,a)-E_{\mathrm{sph}}$ for different lattice volume $L$ and lattice spacing $a$ with $\omega=1/2$. 
    }
    \label{fig:finitesize}
\end{figure}

Substituting the constraints Eq. (\ref{eq:boundaryeta}) into \eq{latt-energy}, we numerically solve for the minimum values and corresponding solutions for different values of $N$ within the domain $\xi_k \in [0,2\pi],~k\in[-N,N)$. These solutions $\{\xi^{\rm sph}_k\}$, together with $\eta_k =\pi/2$, gives the saddle point, i.e., the sphaleron configuration $\{\mathbf{n}^{\rm sph}_k\}$ of the O(3) field on lattice, whose energy is denoted by $E_{\rm sph}(L,a)$. To reliably study sphaleron configurations on the lattice, we need to systematically control the distortions introduced to the sphaleron configurations from the finite volume $L$ and finite lattice spacing $a$. 
We systematically compare the sphaleron energy on the lattice to its continuum value $E_{\rm sph}$ across different lattice volumes $L$ and spacings $a$. The numerical result is shown in FIG.\ref{fig:finitesize}. 

\begin{figure}[h]
\centering
    \includegraphics[width=0.45\textwidth,height=0.3\textwidth]{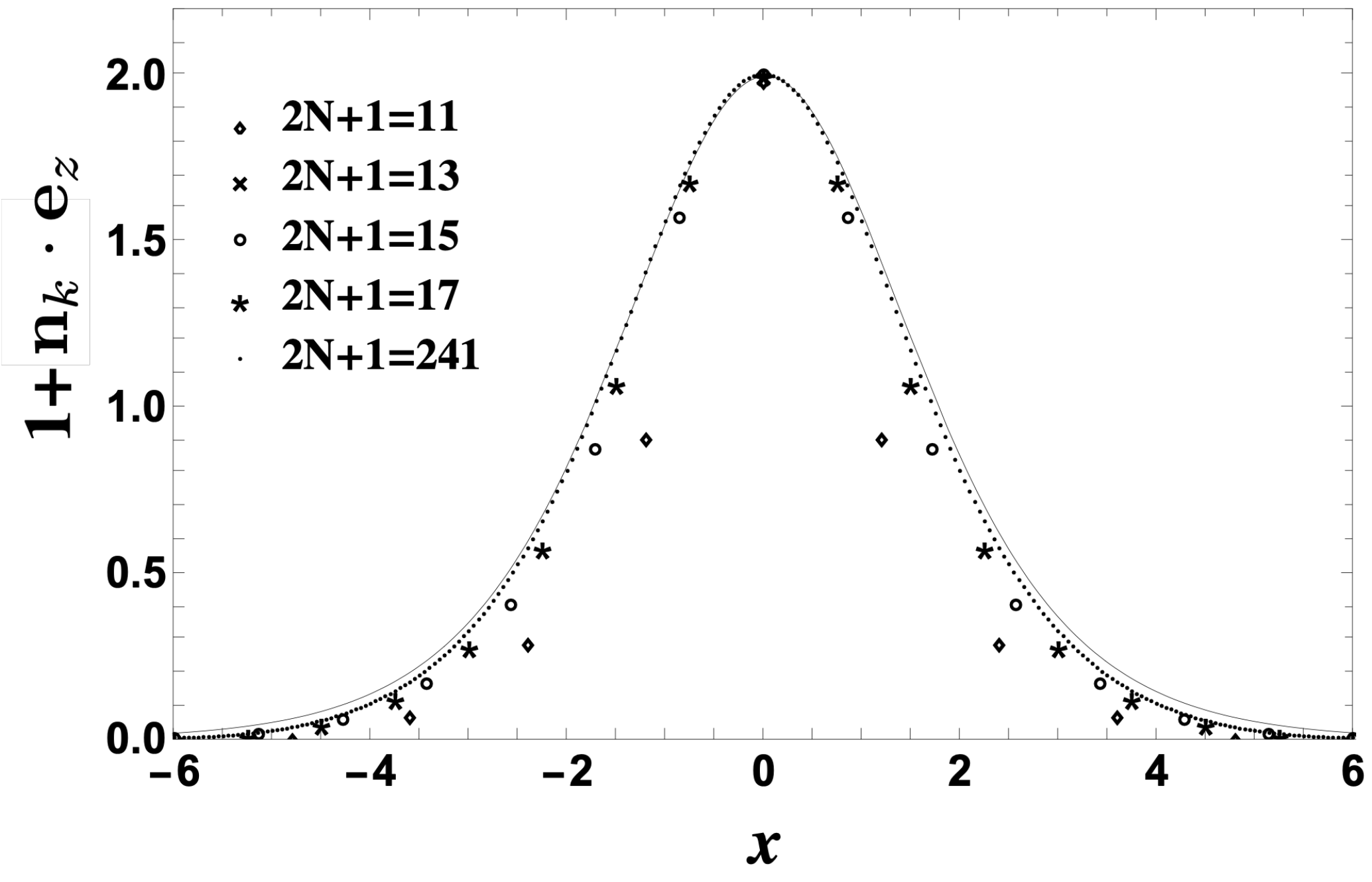}
    \caption{The sphaleron configuration on the lattice across different number of lattice sites. The parameters are chosen to be $L=12$, $\omega=1/2$, and $g=1$. The thin black line is the sphaleron solution in the continuum. 
    }  \label{fig:cosxi}
\end{figure}

\begin{figure}[h]
\centering
    \includegraphics[width=0.47\textwidth,height=0.32\textwidth]{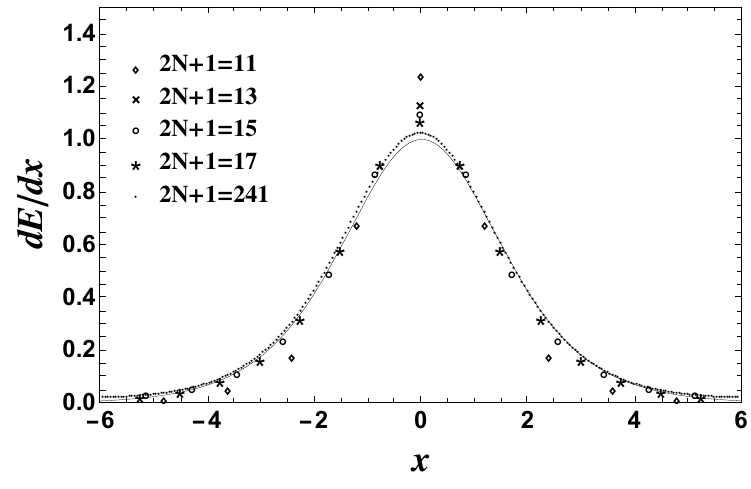}
    \caption{The energy density of the sphaleron configuration on the lattice across different number of lattice sites $2N+1$. The parameters are chosen to be $L=12$, $\omega=1/2$, and $g=1$. The thin black line is the sphaleron solution in the continuous limit. To get a symmetric distribution, we modify the kinetic energy part of the $k$th site in Eq. (\ref{eq:latt-energy}) to be $[1-\frac{1}{2}\cos (\xi_k -
  \xi_{k + 1})-\frac{1}{2}\cos (\xi_{k-1} -
  \xi_{k })]$.}  \label{fig:energydensity}
\end{figure}

To further investigate the convergence to the continuous limit $a\to 0$, we compare the distribution of $1+{\mathbf n}_k \cdot \mathbf{e}_z$ and the energy density $dE/dx$ of sphaleron configuration on the lattice with the continuous one in \fig{cosxi} and \fig{energydensity}. We observe that as the number of lattice sites $2N+1$ increases, \textit{i.e.} the lattice spacing $a$ decreases, the lattice sphaleron configuration approaches the continuum solution. Additionally, a rapid convergence of the lattice sphaleron configuration can be observed as the system size increases to $2N+1 \geq 15$, with errors becoming well-controlled.  

\section{The Quantum Simulations}
\label{sec:4}
To initiate our study of sphaleron production and decay in high-energy collisions, we develop the formulations to simulate the process of sphaleron decay, which require not only defining the classical sphaleron configuration in the framework of quantum field, but also accounting for the quantum effects of the $\mathbf{n}$ fields to incorporate sphaleron decaying to particles. At quantum level, the Hamiltonian describing the dynamics of $\mathbf{n}$ field can be written as
\begin{eqnarray}
    \hat{H}_{\mathrm{tot}} &=& \frac{g^2}{2 a} \sum_{k = -N}^{N}
   \hat{\mathbf{L}}^2_k + \frac{1}{a g^2} \sum_{k = -N}^{N - 1}
   (\mathbb{1} - \hat{{\mathbf n}}_k \cdot \hat{{\mathbf n}}_{k + 1}) \nonumber\\
   && + \frac{a \omega^2}{g^2} \sum_{k = -N}^{N} (\mathbb{1} + \hat{{\mathbf n}}_k \cdot \mathbf{e}_z),
   \label{eq:quant-ham}
\end{eqnarray}
which is lifting the field variable $\mathbf{L}_k$ and ${\bm n}_k$ in $H^{\rm latt}_{\mathrm{tot}}$ as operator $\hat{\mathbf{L}}_k$ and $\hat{\bm n}_k$ acting on local fields at site $k$. The commutation relations follow those of the quantum rotor models, as given by \cite{Sachdev_2011}
\begin{align}
    [L^\alpha, L^\beta] = i \epsilon^{\alpha\beta\gamma}L^
    \gamma, ~~~[L^\alpha, n^\beta]=i\epsilon^{\alpha\beta\gamma} n^\gamma \notag\\
    [n^\alpha, n^\beta] = 0, ~~~\alpha, \beta, \gamma\in\{x, y, z\},
    \label{eq:commutation}
\end{align}
where we have used unbolded notation with the subscript of the lattice index suppressed to denote the individual components of the vector operators $\bold{L}_k$ and $\bold{n}_k$.
The local field can be represented the state $|l, m\rangle_k$ which are eigenstates of the operators $\hat{\mathbf{L}}_k^2$ and $\hat{\mathbf{L}}_k \cdot \mathbf{e}_z$. The state $|l,m\rangle_k$ form the angular momentum basis at local sites. After imposing the cutoff at $l_\text{max}$, $\mathbf{L}_k^2$ becomes a finite-dimensional matrix with eigenvalues $l(l+1)$ for $l = 0,1,\dots,l_{\text{max}}$, where each eigenvalue has multiplicity $2l + 1$ corresponding to the degeneracy 
in the quantum number $m \in \{-l, -l+1, \dots, l\}$. Note that for a finite truncation $l_{\rm max}$, the commutation relation in \eq{commutation} is violated. The dimension of the local Hilbert space at each site becomes $\mathbf{dim} = (l_\text{max}+1)^2$, requiring $n_q = \lceil \log_2(\mathbf{dim}) \rceil$ qubits to represent the local field at a single site when implementing the system on a quantum device. The field degrees of freedom at site $k$ can be mapped to qubits by encoding the states $|l, m\rangle_k$ into the computational basis states of the qubit system. 
The matrix representation of operators $\hat{\mathbf{n}}_k \cdot \mathbf{e}_i$ ($i=x,y,z$) in the computational basis $|l, m\rangle_k$ can be obtained by noting that $\hat{\mathbf{n}}_k \cdot (\mathbf{e}_x \pm i\mathbf{e}_y) = \mp \sqrt{2}\hat{X}_{\pm1}$ and $\hat{\mathbf{n}}_k \cdot \mathbf{e}_z =\hat{X}_0$ with matrix representation of $\hat{X}_M$ given by ~\cite{Bruckmann:2018usp}:
\begin{eqnarray}
\langle l, m \left | \hat{X}_M \right| l', m' \rangle &=& (-1)^m \sqrt{(2l + 1)(2l' + 1)} \times \nonumber\\
&& \begin{pmatrix}
l & l' & 1 \\
0 & 0 & 0
\end{pmatrix}
\begin{pmatrix}
l & 1 & l' \\
-m & M & m'
\end{pmatrix}.
    \label{eq:direction-operator}
\end{eqnarray}

The real-time evolution for $|\psi_{\rm init}\rangle$ under $\hat{H}_{\mathrm{tot}}$ can be realized using Trotterization scheme~\cite{Suzuki:2002zuq}. Consider the three non-commuting terms in $\hat{H}_{\rm tot}$ as 
\begin{align}
    \hat{H}_1 &= \frac{g^2}{2 a} \sum_k \hat{\mathbf{L}}_k^2, \notag\\
    \hat{H}_{\rm 2} &= \frac{1}{a g^2} \sum_k (\mathbb{1} - \hat{\mathbf{n}}_k \cdot \hat{\mathbf{n}}_{k+1}), \notag\\
    \hat{H}_{\rm 3} &= \frac{a \omega^2}{g^2} \sum_k (\mathbb{1} + \hat{\mathbf{n}}_k \cdot \mathbf{e}_z),
    \label{eq:Hamdecompose}
\end{align}
the second-order Trotterization for evolving $\hat{H}_{\mathrm{tot}}$ for a total time $t$ with trotter step $\Delta t$ can therefore be written as $\left[U(\Delta t)\right]^s$, with $s = t/\Delta t$ and
\begin{align}
    U(\Delta t) = \prod^{l=3}_{l=1}U_l(\Delta t)\prod^{l=0}_{l=3}U_l(\Delta t).
    \label{eq:trotter}
\end{align}
with $U_l(\Delta t)=e^{i \hat{H}_l \Delta t/2}$.

In the following, we describe how to implement the time evolution operator in \eq{trotter} using quantum circuits. When an estimation of the circuit depth is necessary, we focus on the universal quantum gate set $\{R_x(\theta), R_y(\theta), R_z(\theta), {\rm CNOT}\}$ where the gates included are widely supported by near-term devices. Given that operators acting on different lattice sites commute and $\hat{H}_{1}$ is diagonal in the computational basis, $e^{i \hat{H}_l \Delta t/2}$ can be implemented with $R_Z(\theta)$ gates and CNOT gates with a circuit depth of $\mathcal{O}(2^{n_q})$ for general cases. Define $U_{k,k+1}(\Delta t) = \exp(-i \theta_1\Delta t \hat{\mathbf{n}}_k \cdot \hat{\mathbf{n}}_{k+1})$ and $U_{k,z}(\Delta t) = \exp(-i \theta_2\Delta t\hat{\mathbf{n}}_k \cdot \mathbf{e}_z)$ with $\theta_1 = 1/2ag^2$ and $\theta_2 = a\omega^2/2g^2$, we have
\begin{eqnarray}
    &&U_2(\Delta t) = e^{i\theta_1 \Delta t}\prod_k U_{k,k+1}(\Delta t), \notag\\
    &&U_3(\Delta t) =e^{i\theta_2 \Delta t} \prod_k U_{k,z}(\Delta t).
    \label{eq:u23decompose}
\end{eqnarray} 
Given the matrix representation of $\hat{\mathbf{n}}_k \cdot \mathbf{e}_i$ in the computational basis, $\hat{\mathbf{n}}_k \cdot \hat{\mathbf{n}}_{k+1}$ and $\hat{\mathbf{n}}_k \cdot \mathbf{e}_z$ can be systematically decomposed into linear combinations of Pauli strings with operator-Schmidt techniques \cite{nielsen_chuang_2010}, allowing the time evolution operator $U_{k, k+1}(\Delta t)$ and $U_{k, z}(\Delta t)$ to be implemented with quantum circuits using trotterizations or qDRIFT algorithm \cite{qDrift}. For the case of $n_q <12$ where matrix exponentiation is classically tractable, these evolution operators can also be exactly decomposed into quantum gates. The time evolution operators $U_2(\Delta t)$ and $U_3(\Delta t)$ can thus be constructed subsequently following \eq{u23decompose}.

For the quantum algorithm based on a lattice with $2N+1 = 5$ sites and the simplest nontrivial truncation $l_{\rm max} =1$ which lead to $n_q = 2$s. In this case, the mapping of the local states to the computational basis can be chosen as $|l=0, m=0\rangle_k\rightarrow  |00\rangle_k$,  $|l=1,m=0\rangle_k\rightarrow  |01\rangle_k$, $|1-1\rangle_k\rightarrow  |l=1,m=-1\rangle_k$ and $|l=1, m=1\rangle_k\rightarrow  |11\rangle_k$ at site $k$. Here, $|00\rangle_k$, $|01\rangle_k$, $|10\rangle_k$ and $|11\rangle_k$ represent the computational basis states of the two qubits at lattice site $k$. $\hat{\mathbf{n}}_k \cdot \mathbf{e}_z$ can be decomposed as a summation of $K_1=4$ independent Pauli strings, and $\hat{\mathbf{n}}_k \cdot \hat{\mathbf{n}}_{k+1}$ as a summation of $K_2=34$ independent Pauli strings, where most of the Pauli strings are non-commuting with each other. A general Pauli string of dimension $2^{n_q} = 16$ requires a CNOT circuit depth of $\mathcal{O}(50)$ for implementing its corresponding time evolution operators. The error $\epsilon_{ps}$ of implementing $U_{k, k+1}(\Delta t)$ and $U_{k, z}(\Delta t)$ using $r$ trotter steps can be estimated as $\mathcal{O}(2N K_2\Delta t^2/a^2 g^4 r)$ and $\sim \mathcal{O}(2N K_2\Delta t^2 a^2 \omega^2/g^4 r)$, respectively for a first order trotterization algorithm. For $\epsilon_{ps}\sim 1\%$, $\Delta t/a \sim 0.1$, and $g =1$, we obtain $r\sim 100$, which leads to a circuit depth of $\mathcal{O}(5\times 10^3)$ for the case of $U_{k, k+1}(\Delta t)$. For comparison, the exact decomposition of $U_{k, k+1}(\Delta t)$ using \textsc{Qiskit} transpiler gives a CNOT circuit depth of $\mathcal{O}(3\times 10^4)$. Given the substantial circuit depths involved, implementing the time evolution on a NISQ device would require dedicated quantum error mitigation/correction to obtain reliable results, which we leave for future investigations. 

We expect to study the evolution of sphalerons within the above framework and extract physical observables, such as sphaleron lifetime, sphaleron decay width to particles. Yet, how to define a quantum state that can be the correspondence of the classical configuration remains elusive and of vital importance. In the following, we layout two possible directions to be explored for this purpose. One attempt would be constructing the sphaleron initial states as quantum coherent state, which can interpolate between quantum and classical descriptions \cite{RevModPhys.62.867}. Quantum coherent states corresponding to the classical field configurations of a sphaleron have been used to study the decay products of sphalerons \cite{ZADROZNY199288} in the framework of SM. However, within the framework of the nonlinear $O(3)$ $\sigma$-model, and in particular under the angular momentum quantization scheme defined by the Hamiltonian in \eqref{eq:quant-ham}, the standard construction of coherent states, i.e. Glauber-type coherent states~\cite{PhysRev.131.2766}, associated with the Weyl--Heisenberg algebra and particle-number quantization is not applicable. As Perelomov emphasized~\cite{Perelomov:1971bd, Perelomov:1986uhd}, the Glauber-type coherent states rely essentially on the nilpotent and non-compact group structure of the Heisenberg--Weyl group, and cannot be straightforwardly extended to compact groups such as $O(3)$. For this reason, 
 
One may lead to consider the group-theoretical framework introduced by Perelomov, in which generalized coherent states are defined by acting with the group transformations on a chosen reference state in the Hilbert space. However, the choice of reference state and group action is not unique. The influence of this ambiguity on the $O(3)$ sphaleron process remains an open question, necessitating further investigations.

Another attempt is to consider the classical sphaleron configuration as a background field. This strategy has been used in \cite{HELLMUND1992749} to study the classical evolution of sphalerons in the SM and calculate the averaged number of Higgses and W-bosons from sphaleron decays. To allow interpretations of particles produced exclusively, quantizations are necessary. Yet, such classical evolution of sphalerons might be taken as background fields added to the quantized Hamiltonian to allow studies of partial decay width, particle number distributions etc. For our case in the $O(3)$ model, it would result in the following Hamiltonian by neglecting the constant terms in \eq{quant-ham}:
\begin{eqnarray}
    &&\hat{H}^{\rm bg}_{\mathrm{tot}} = - \frac{1}{a g^2} \sum_{k = -N}^{N - 1}
   \bigg(\hat{{\mathbf n}}_k + \mathbf{n}_k(t)\bigg) \cdot \bigg(\hat{{\mathbf n}}_{k + 1}+ \mathbf{n}_{k+1}(t)\bigg)\notag\\
   &&~~~~~ + \frac{a \omega^2}{g^2} \sum_{k = -N}^{N}  \bigg(\hat{{\mathbf n}}_k+ \mathbf{n}_k(t)\bigg) \cdot \mathbf{e}_z +\frac{g^2}{2 a} \sum_{k = -N}^{N}
   \hat{\mathbf{L}}^2_k.
   \label{eq:quant-ham}
\end{eqnarray}
In the above, we have introduced $\mathbf{n}_k(t)$ as the time-dependent classical background field with $\mathbf{n}_k(t= 0)\equiv \mathbf{n}^{\rm sph}_k$. This method, with the classically computed $\mathbf{n}_k(t)$ driving $\hat{H}^{\rm bg}_{\mathrm{tot}}$, should offer a clear path to study the exclusive decay of sphalerons into particles, a direction we pursue in future work.

\section{Conclusions}
\label{sec:5}
The production of sphaleron at high energy collisions in the Standard model is yet to be observed. Despite decades of theoretical investigation, reliable predictions of sphaleron production rates remain elusive. As a nonperturbative quantum process requiring real-time field theory simulations, this challenge may ultimately be resolved through quantum computation. In this work, we initiate this research program by studying the $1+1$ dimensional $O(3)$ model, which, when coupled to fermions, provides a tractable framework for studying sphaleron production in high-energy collisions within electroweak theory. With lattice formulation, we explicitly identified the sphaleron configuration as compared to that in the continuous spacetime theory. Furthermore, we established appropriate lattice volumes and spacings that reproduce the sphaleron energy of the continuum theory with controlled precision. To simulate dynamical processes involving sphaleron production from particle collisions and subsequent decay, we have developed the quantum algorithms for the sphaleron evolution using trotterization. Future work should extend these efforts along three directions: developing quantum algorithms to extract sphaleron production rates in high-energy collisions as well as its decay branching ratio to particle, measuring Chern-Simons number evolution during sphaleron dynamics, and incorporating fermionic degrees of freedom into the framework.
\begin{acknowledgments}
The work of M. H and H. Z is supported in part by the National Science Foundation of China under Grants No. 12075257 and No. 12235001. Particularly, we would like to thank Dr. K. Hu at the Particle Astrophysics Center of the Institute of High Energy Physics (IHEP), Chinese Academy of Sciences, for his support and assistance in the programming of quantum simulations. Y.-Y. L is supported by IHEP under Grant No. E55153U1. 
\end{acknowledgments}

\bibliography{main}

\appendix
\section{Constructing the discretized Hamiltonian}
\label{sec:app}
Rewrite the phase space coordinates of ${\mathbf n}$ using polar coordinates.
\begin{eqnarray*}
{\mathbf n}(\theta, \varphi) & = & (\sin \theta (x) \cos \varphi (x), \sin \theta
   (x) \sin \varphi (x), \cos \theta (x)),
\end{eqnarray*}
the Lagrangian can be written as
\begin{eqnarray*}
\mathcal{L} & = & \frac{1}{2 g^2} (\partial_{\mu} {\mathbf n})^2 -\frac{\omega^2}{g^2}(1+{\mathbf n}\cdot\vec{e}_z)\\
  & = & \frac{1}{2 g^2} (\dot{\theta}^2 + \dot{\varphi}^2 \sin^2 \theta -
  (\partial_x {\mathbf n})^2)-\frac{\omega^2}{g^2}(1+{\mathbf n}\cdot\mathbf{e}_z)
\end{eqnarray*}
With
\begin{eqnarray*}
\frac{\partial \mathcal{L}}{\partial \dot{\theta}} & = & \frac{\dot{\theta}^2}{g^2} \nonumber\\
\frac{\partial \mathcal{L}}{\partial \dot{\varphi}} & = & \frac{\dot{\varphi}^2
   \sin^2 \theta}{g^2} \nonumber\\
\end{eqnarray*}
the classical Hamiltonian can be obtained via Legendre transformation as:
\begin{eqnarray}
H_{\rm tot} & = & \frac{1}{2g^2}\int dx\bigg(\dot{\theta}^2 + \dot{\varphi}^2 \sin^2
   \theta + (\partial_x {\mathbf n})^2\nonumber\\
   &&+2\omega^2(1+{\mathbf n}\cdot\mathbf{e}_z)\bigg).
\end{eqnarray}
where the first two terms are the kinetic energy, and the rest are taken as potential energies. 

Next, we proceed with discretization in the $x$-direction while omitting
the constant term:
\begin{eqnarray}
H^{\rm latt}_{\rm tot} &=& \frac{a}{2 g^2} \underset{k}{\overset{}{\sum}} \left( \dot{\theta}_k^2 + \dot{\varphi}_k^2 \sin^2 \theta_k - \frac{2}{a^2} {\mathbf n}_k
   {\mathbf n}_{k + 1} \right)\\&&+ \frac{a \omega^2}{g^2} \underset{k}{\overset{}{\sum}} (1+{\mathbf n}_k\cdot{\mathbf{ e}_z}) \notag\\
   &=& \underset{k}{\overset{}{\sum}} \left(\frac{1}{2} I {{\bm \omega}}_k^2 - \frac{1}{2 g^2} \frac{2}{a} {\mathbf n}_k
   {\mathbf n}_{k + 1}\right) \notag\\&&+ \frac{a \omega^2}{g^2} \underset{k}{\overset{}{\sum}} (1+{\mathbf n}_k\cdot{\mathbf{ e}_z}).
\end{eqnarray}
Since the phase space is a spherical surface, the
kinetic energy part is expressed as rotational energies in terms with the moment of inertia $I=a/g^2$ and the angular velocity vector
with the angular velocity vector:
\begin{align} 
\nonumber {{\bm \omega}}_k &= \begin{pmatrix}
\cos \theta_k \cos \varphi_k \dot{\theta}_k - \sin \varphi_k  \sin \theta_k  \dot{\varphi}_k  \\
\cos \theta_k \sin \varphi_k \dot{\theta}_k + \cos \varphi_k  \sin \theta_k  \dot{\varphi}_k \\
- \sin \theta_k \, \dot{\theta}_k
\end{pmatrix}.
\end{align}
We thus define the corresponding angular
momentum as
\begin{align}
{\mathbf L}_k (\theta_k, \varphi_k) \equiv I {\bm \omega}_k
    \label{eq:angular-momentum}
\end{align}
Rewriting the Hamiltonian in terms of angular
momentum in vector form, we obtain:
\begin{eqnarray}
  H^{\rm latt}_{\rm tot} &=& \frac{g^2}{2 a} \underset{k}{\overset{}{\sum}} {\mathbf L}^2_k -
  \frac{1}{ag^2} \underset{k}{\overset{}{\sum}}{\mathbf n}_k {\mathbf n}_{k
  + 1}\notag\\
  &&+ \frac{a \omega^2}{g^2} \underset{k}{\overset{}{\sum}} (1+{\mathbf n}_k\cdot{\mathbf{ e}_z})
\end{eqnarray}

\end{document}